\documentclass[useAMS,usenatbib]{mn2e}
\usepackage{graphicx}
\usepackage{natbib}  
\newcommand\simgt{\lower.5ex\hbox{$\; \buildrel > \over \sim \;$}} 
\newcommand\simlt{\lower.5ex\hbox{$\; \buildrel < \over \sim \;$}}

\newcommand{\msun}{\ensuremath{\, {M}_\odot}}
\newcommand{\Msun}{\ensuremath{\, {M}_\odot}}

\newcommand{\tbce}{T_{bce}}

\title[Self--enrichment by AGB in high metallicity Globular Clusters]{Self--enrichment 
by AGB stars in Globular Clusters: comparison between intermediate and high 
metallicities}
\author[P. Ventura and F. D'Antona]{P. Ventura\thanks{E-mail:
ventura@oa-roma.inaf.it (PV); dantona@oa-roma.inaf.it (FD)} and F. D'Antona
\\
INAF, Osservatorio di Roma, MONTEPORZIO, I-00040, Italy
}
\begin{document}

\date{Accepted . Received ; in original form }

\pagerange{\pageref{firstpage}--\pageref{lastpage}} \pubyear{2007}

\maketitle

\label{firstpage}

\begin{abstract} 
We present theoretical evolutionary sequences of intermediate mass stars 
(M=3-6.5$\msun$) with metallicity Z=0.004. Our goal is to test whether the 
self-enrichment scenario by massive Asymptotic Giant Branch stars
may work for the high metallicity Globular Clusters, after previous works 
by the same group showed that the theoretical yields by this class of
objects can reproduce the observed trends among the abundances of some
elements, namely the O-Al and O-Na anticorrelations, at intermediate 
metallicities, i.e [Fe/H]=-1.3. 
We find that the increase in the metallicity 
favours only a modest decrease of the luminosity and the 
temperature at the bottom of the envelope for the same core mass, and
also the efficiency of the third dredge-up is scarcely altered.
On the contrary, differences are found in the yields,
due to the different impact that processes with the same efficiency have on
the overall abundance of envelopes with different metallicities. We expect
the same qualitative patterns as in the intermediate metallicity case, but
the slopes of some of the relationships among the abundances of some elements
are different.
We compare the sodium--oxygen anticorrelation for clusters of intermediate metallicity
(Z$\approx 10^{-3}$) and clusters of metallicity large as in these new models. Although
the observational data are still too scarce, the models are consistent with the
observed trends, provided that only stars of M$\simgt$5\msun\ contribute to
self--enrichment.

\end{abstract}


\section{Introduction}
Spectroscopic investigations of Globular Clusters (GC) show
star to star differences in their surface chemistries (Kraft 1994). 
These inhomogeneities trace clear abundance patterns involving all the light 
elements: sodium is correlated to aluminium and anticorrelated to oxygen, 
whereas magnesium is anticorrelated with aluminium (Carretta 2006), 
though in some clusters 
the existence of this latter relationship is still under debate 
(Cohen \& Mel\'endez 2005). A common result of these analysis is the approximate 
constancy of the overall CNO abundances (Ivans et al. 1999).

The detection of such anomalies in scarcely evolved stars,
like Turn-Off (TO) and Sub-Giant Branch (SGB) stars (Gratton et al 2001), 
ruled out the possibility of any in situ mechanism as a possible unique 
explanation (Denissenkov \& Weiss 2004),
and pointed in favour of a self-enrichment scenario, i.e. that the
stars with the anomalous chemistry formed from the ashes of a
previous stellar generation, which polluted the interstellar medium
with matter which had been processed via the CNO cycle. Ventura et
al. (2001) indicated the massive Asymptotic Giant Branch (AGB) stars
as likely candidates, because with appropriate hypothesis concerning
convection modelling they can easily achieve Hot Bottom Burning (HBB),
i.e. the bottom of their envelope becomes so hot to ignite strong
nucleosynthesis, whose products, convected to the surface of the star,
are ejected into the interstellar medium via the strong winds suffered
by these sources (Ventura \& D'Antona 2005); also, strong HBB favours 
high mass loss, reduces
the number of thermal pulses (TP) experienced by the stars during their
life, thus diminishing the number of third dregde-up (TDU) episodes,
and keeping approximately constant the C+N+O mass fraction in the
envelope, for a reasonably large range of initial masses.
This conclusion is at odds with the results by Fenner et al. (2004),
who, based on AGB models calculated with the traditional, low efficiency,
mixing length theory (MLT, Vitense 1953) convective model, found that
oxygen can be hardly depleted at the surface of these stars, and the 
CNO sum is expected to largely exceed unity.

Recently, an alternative self-enrichment scenario for GCs was
proposed by Maeder \& Meynet (2006), Prantzos \& Charbonnel (2006),
and described in details by Decressin et al. (2007): in this case
the enrichment of the interstellar medium is produced by the envelopes
of fast rotating massive stars. In order to choose between the AGB and
the massive stars self-enrichment, it is necessary to explore in detail
both models. In particular, the variation of predictions of yields
with metal abundance must be examined.
Ventura \& D'Antona (2008, hereinafter VD08) showed that the
enrichment by AGBs can work in the case of intermediate metallicity GCs, since 
the ejecta of their theoretical models, calculated with a metallicity
Z=0.001, with initial masses in the range 5-6$\msun$, are in qualitative 
and quantitative agreement with the abundance patterns observed in
scarcely evolved stars of M3, M13, M5, NGC6218, NGC6752. 

The goal of the present paper is to investigate whether the self-enrichment
scenario hypothesis may work also in the case of metal rich GCs,
having metallicities [Fe/H]$\sim -0.7$.
We present a new set of AGB models with metallicity Z=4$\times 10^{-3}$, 
typical of metal rich GCs like 47Tuc and M71. We discuss the possibility
of reproducing the O-Al and O-Na anticorrelations, and determine the 
expected slope of these relationships compared to the intermediate
metallicity case. 

\begin{figure}
  \includegraphics[width=.48\textwidth]{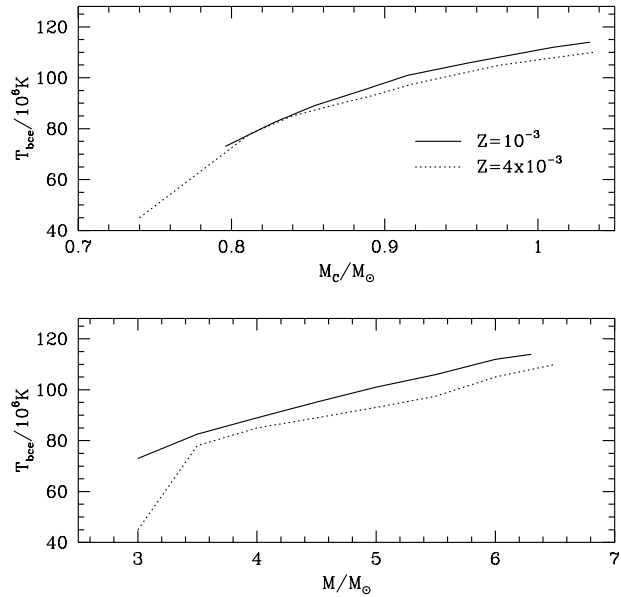}
  \caption{The variation of the temperature at the bottom of the
   convective envelope as a function of the core mass at the
   beginning of the AGB phase (Top) and of the initial mass (Bottom)
   of AGB models with two different metallicities. The temperatures
   refer to the phase of maximum luminosity during the AGB evolution.}
  \label{phys}
\end{figure}

\section{The models}
\subsection{Physical and chemical inputs}
The evolutionary sequences were calculated by means of the ATON code
for stellar evolution, a full description of which is given in Ventura
et al. (1998), with the last updates given in Ventura \& D'Antona (2005).
The physical and chemical inputs coincide with those given in VD08,
with the only exception of the metallicity, for which we use 
Z=4$\times 10^{-3}$ in the present investigation. We briefly recall
the main assumptions:
\begin{itemize}
\item{Convection is modelled according to the Full Spectrum of Turbulence
(FST) model by Canuto \& Mazzitelli (1991)}

\item{Mass loss is treated according to Bl\"ocker (1995); for the free
parameter entering the Reimer's formula, we used $\eta_R=0.02$}

\item{Nuclear burning and mixing of chemicals are treated simultaneously,
following the diffusive approach by Cloutmann \& Eoll (1976); the parameter
$\zeta$ giving the exponential decay of convective velocities within regions
radiatively stable was set to $\zeta=0.02$. The debate concerning a 
possible extra-mixing from the bottom of the convective envelope during
the TDU is still open, and the uncertainties associated to this issue are
such that we prefer not to include it in the present work (see Herwig 2005
for a exhaustive discussion on this topic).}

\item{The NACRE compilation was used for all the relevant cross-sections,
with the only exception of the Ne-Na proton capture rates, for which we
used the formulae given by Hale et al. (2002; 2004), and the nitrogen
proton capture cross section, for which we used the new rate given by
Formicola et al. (2004)}

\item{The mixture is assumed to be $\alpha$-enhanced, with $[\alpha$/Fe]=+0.4
\footnote{We use the standard notation, defining 
[X/Fe]=$\log(X/Fe)-\log(X/Fe)_{\odot}$}. Compared to the standard
solar mixture, the abundances of $^{16}$O, $^{20}$Ne and $^{24}$Mg are
increased. The individual abundances are taken from Grevesse \& Sauval (1998)}

\end{itemize}

\subsection{The physical properties}
All the models were followed from the pre-MS phase up to the almost
complete consumption of the convective envelope. We find HBB in all cases,
with the only exception of the 3$\msun$ model.

Fig.\ref{phys} shows the comparison between the main physical properties
of the present set of models and those published in VD08. The two panels
show the temperature at the bottom of the envelope ($\tbce$) as a
function of the initial mass of the star (bottom panel) and of the core mass
($M_C$)at the beginning of the AGB phase (top). The plotted temperatures
refer to the phase when the maximum luminosity is reached,
before the decline of log$(L/L_{\odot})$ due to the 
reduction of the mass of the envelope: these quantities can be used 
as indicators of the degree of nucleosynthesis which we expect at 
the bottom of the outer convective zone.

As expected, for a given initial mass, the lower metallicity models
are hotter and more luminous; yet, the $M_C-\tbce$
relation is approximately the same for the two set of models, 
thus the larger metallicity simply increases the initial
mass at which a given degree of nucleosynthesis can be achieved in the
outer convective envelope.
The maximum $\tbce$ reached in the present set of models is
$\tbce=107$MK, i.e. almost coincident with the maximum value
found in VD08.
The efficiency of the third dredge up is also similar to what
was found in the lower metallicity case: the parameter\footnote{We 
indicate with $\lambda$ the ratio between the
inner penetration (in mass) of the bottom of the convective envelope during 
the TDU and the advance of the CNO shell from the previous TP}
$\lambda$ is $\lambda=0.3$ during the latest evolutionary phases of 
the 6$\msun$ model, and increases up to 0.7 for $M=4\msun$. 

\subsection{The chemical yields}
Table 1 reports the average chemical content of the ejecta of the 
$Z=4\times 10^{-3}$ set of models.

%
\begin{table*}
\caption{Chemical composition of the ejecta of intermediate mass models}             
\label{tabchim}      
\centering          
\begin{tabular}{c c c c c c c c c c c c}     
\hline\hline       
$M/M_{\odot}$ & Y & Li & [$^{12}$C/Fe] & [$^{14}$N/Fe] &
[$^{16}$O/Fe] & [$^{23}$Na/Fe] & [Mg/Fe] & [$^{27}$Al/Fe] & 
R(CNO) \\ 
\hline                    
   3.0 & .277  & 1.42  &  1.41   & 0.60  &  0.62  &   0.44  &  0.59  &  0.68  & 4.7    \\  
   3.5 & .269  & 2.63  &  0.08   & 1.59  &  0.46  &   1.07  &  0.50  &  0.33  & 2.5    \\
   4.0 & .281  & 2.20  & -0.07   & 1.52  &  0.30  &   1.17  &  0.48  &  0.32  & 2.0    \\
   4.5 & .298  & 2.00  & -0.44   & 1.52  &  0.21  &   1.00  &  0.47  &  0.43  & 1.8    \\
   5.0 & .313  & 1.98  & -0.55   & 1.44  &  0.09  &   0.89  &  0.45  &  0.57  & 1.4    \\
   5.5 & .328  & 1.93  & -0.62   & 1.37  &  0.01  &   0.76  &  0.43  &  0.70  & 1.2    \\
   6.0 & .329  & 2.02  & -0.78   & 1.25  &  0.01  &   0.63  &  0.42  &  0.71  & 1.0    \\
   6.5 & .330  & 2.06  & -0.85   & 1.19  &  0.05  &   0.60  &  0.43  &  0.66  & 0.96   \\
\hline                 
\end{tabular}
\end{table*}

We note systematic differences compared to VD08, particularly
for the most massive models: aluminum is produced at a lower extent, oxygen is
less depleted, and sodium is synthesized in greater quantities. Also, the
CNO ratio, giving the increase of the overall CNO abundances with respect to
the initial mass fractions, exceeds 2 only for the lowest masses.
The carbon and nitrogen yields are also different: the $^{12}C$ yields are
systematically lower, whereas the nitrogen production is similar in the two
cases.

\begin{figure*}
\includegraphics[width=.48\textwidth]{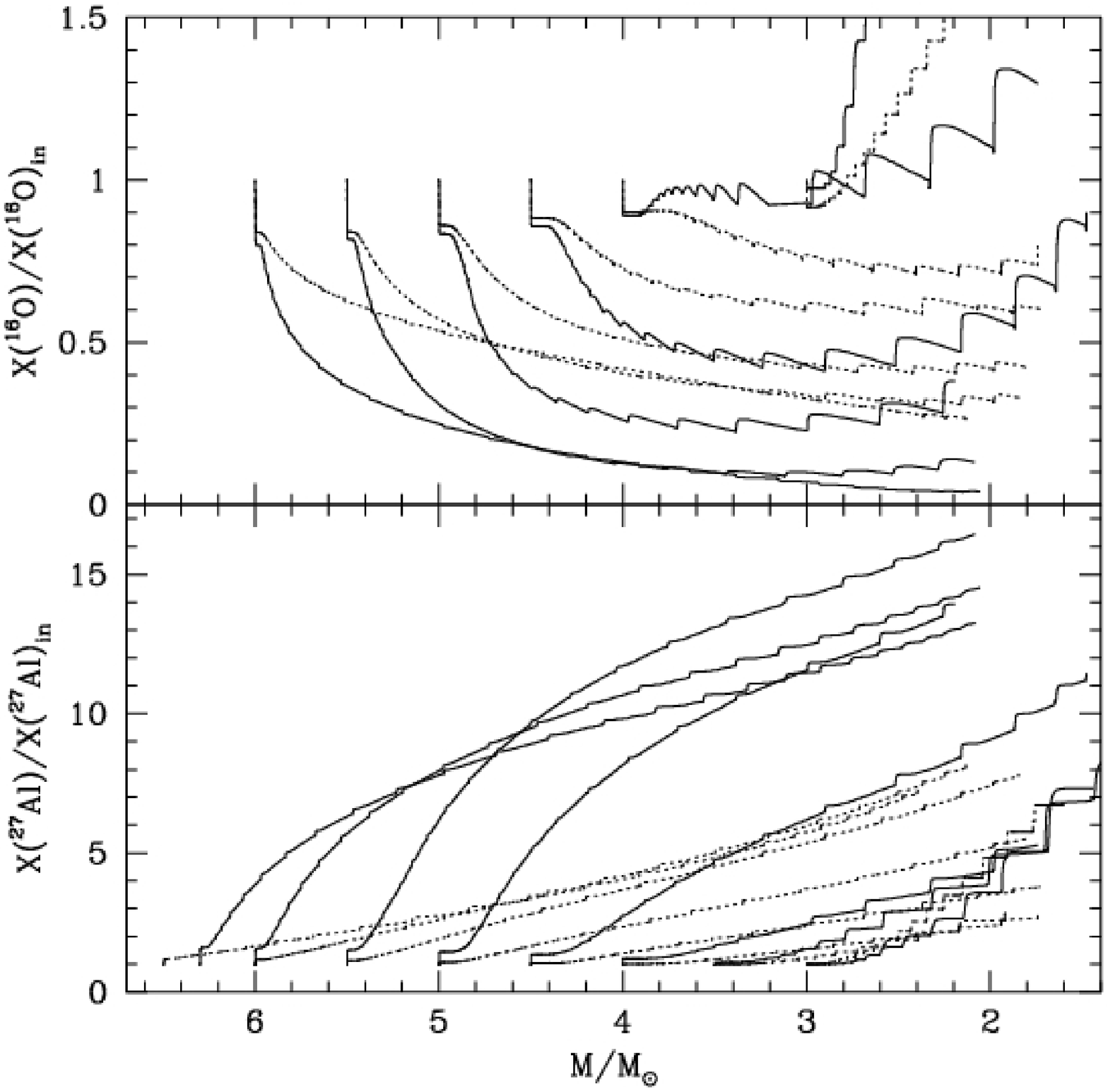}
\includegraphics[width=.48\textwidth]{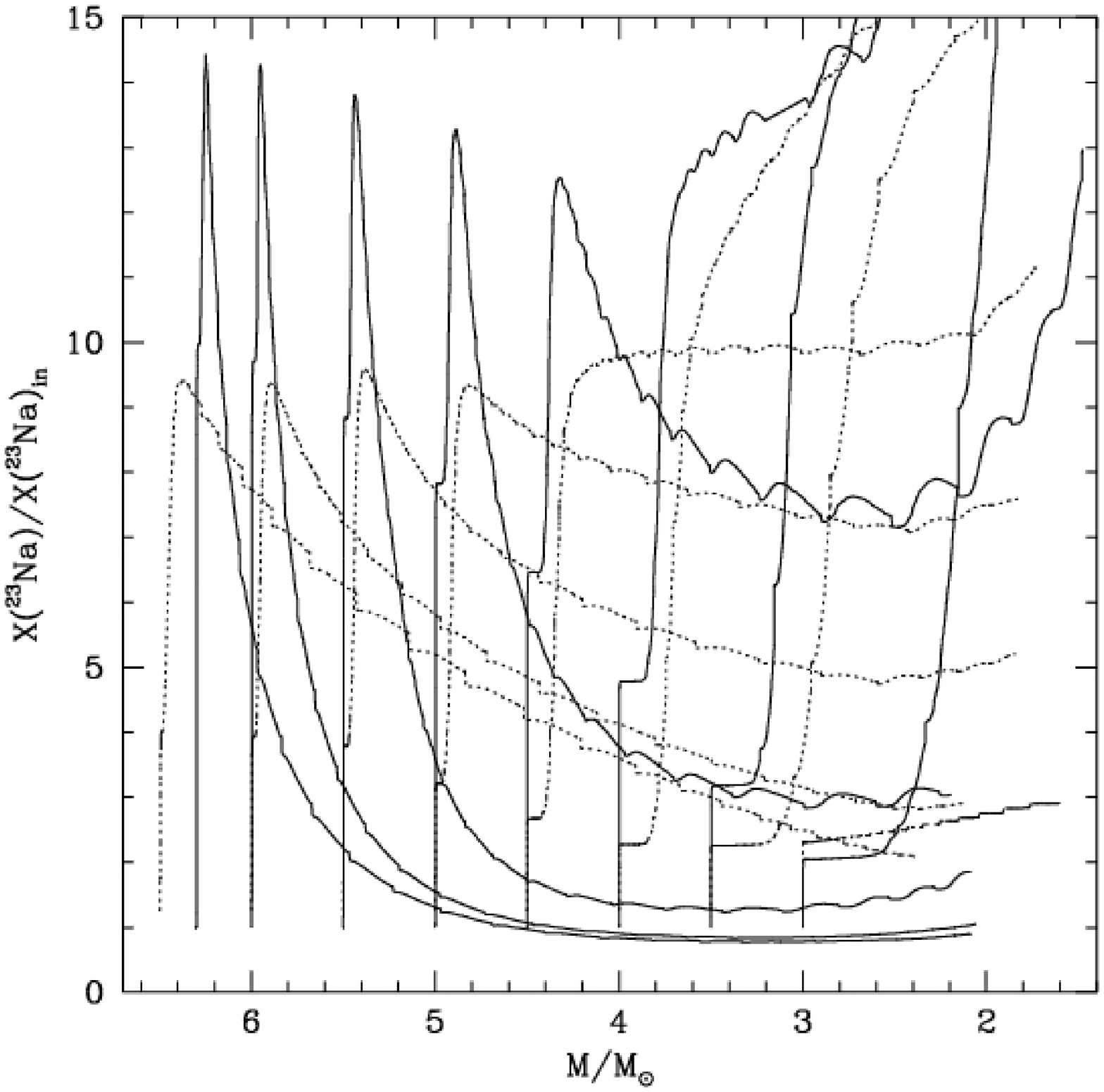}
\vspace{-0pt}
\caption{{\bf Left:} The variation with the total stellar mass of
 the surface abundances of oxygen (Top) and aluminium (Bottom) during
 the AGB phase of intermediate mass models with metallicity $Z=10^{-3}$
 (solid track) and  $Z=4\times 10^{-3}$ (dotted). For both elements, the
 ratio with respect to the initial values are shown. {\bf Right:} The
 variation of the surface sodium abundances for the same models of the
 left panel.}
 \label{OAl}
\end{figure*}

The aforementioned differences cannot be ascribed to the different
nucleosynthesis achieved at the bottom of the convective zone, since we
find similar temperatures for the same core mass, independently of Z.
The main point here is that increasing the overall metallicity, hence
the mass fractions of the individual species, makes harder to change the
abundances of the elements in the whole envelope: this holds both in
the cases of HBB and TDU. Particularly for this latter mechanism, 
less relevant changes are expected at higher metallicity for the same 
$\lambda$.

To make this point clearer , we show in the left panel of Fig.\ref{OAl} the 
variation of the abundances of oxygen (top) and aluminium (bottom) for the 
two sets of models; the right panel shows the behaviour of the surface
sodium. We chose the mass as abscissa, to have a direct idea of the yield 
expected for a given element, whereas on the y-axis we put the ratio of the 
current over the initial abundance, thus providing an indication of the 
change of the individual mass fractions.

It is clear from the left panel of Fig.\ref{OAl} that for both O and Al the changes
of the surface abundances are smaller in the Z=0.004 set. Oxygen is depleted
at approximately the same extent as in the Z=0.001 case, yet the variation
with respect to the starting abundance is smaller (note that this trend is
opposite for the lowest masses, where oxygen is indeed produced by the TDU,
and never burnt, due to the modest HBB found for $M\leq 4\msun$). 
The dichotomy in the
Al production is even more evident: Aluminum is produced both via HBB and
also, indirectly, by the TDU, and both mechanisms, for the same temperatures
and $\lambda$, have a modest effect on the surface chemistry in the Z=0.004
case. With respect to sodium (see the right panel of Fig.\ref{OAl}), the 
question is more tricky, since the trend with mass changes during the evolution.
We note a larger percentage increase in the low metallicity models at the
beginning of the AGB evolution, but a stronger depletion in the following
phases, when the proton capture reactions by $^{23}$Na nuclei favours the
sodium depletion; since this latter phase is longer than the production
process, we find on the average a larger production of sodium in the Z=0.004
models.

\subsection{A comparison with other AGB yields}
We compare our AGB yields with those found, for the same range of 
masses and metallicity, with the most extended, published, computations,
based on real model evolution, by Karakas \& Lattanzio (2007, hereinafter KL07).

\begin{figure}
\includegraphics[width=.48\textwidth]{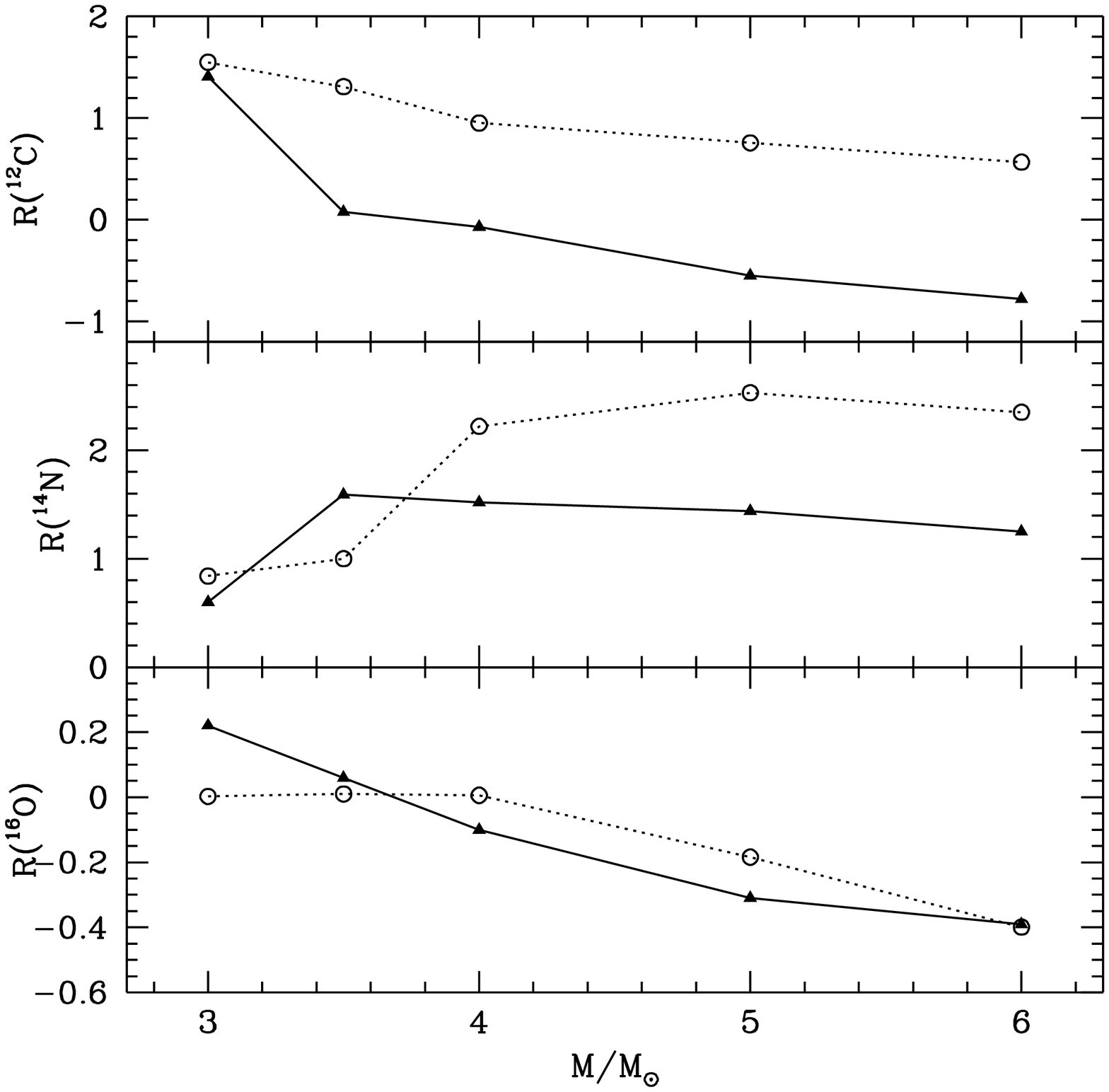}
\caption{The average mass fractions of carbon (top panel), nitrogen
(medium) and oxygen (bottom) in the ejecta of AGB models as a
function of the initial mass of the star. For each element i, we show
in the y-axis the quantity Log($X_i/X_i^0$), where $X_i^0$ is the
initial abundance. The full triangles and the open circles indicate,
respectively, the results from the present investigation and those
from the KL07 models.}
\label{amanda}
\end{figure}

The comparison between the results is shown in Fig.~\ref{amanda}. 
The three panels show, for each value of the initial mass, the CNO abundances; 
the quantity
shown in the y-axis is actually the ratio of the average abundance
compared to the initial value, in a logarithmic scale.

For the largest masses of our sample, we see two important differences
between the two sets of models:

\begin{enumerate}

\item{Our models destroy carbon, up to 1dex for the 6$\msun$ model,
while the models by KL07 actually produce it in all
cases}

\item{Both sets of models produce nitrogen, but our models production
is a factor $\sim 10$ smaller than KL07}

\item{The CNO sum is very different in the two cases: it is approximately
constant in our simulations (see col.10 of tab.~\ref{tabchim}), whereas
it increase by a factor 10 in the KL07 models}

\end{enumerate}

These differences, though large, are actually not surprising, since
the two sets of models have been calculated with a different treatment
of convection and of mass loss. Ventura \& D'Antona (2005a) showed
that when convection is modelled efficiently, e.g. with the FST
model used in the present investigation, HBB is favoured, the evolution
is much faster, and the number of TPs and of TDU episodes experienced
by the stars during the AGB phase is greatly reduced.
The much larger abundances of carbon and nitrogen found in the
KL07 are a mere consequence of the many TDUs experienced by their
models, as a direct consequence of the much higher number of TPs
determined by the MLT modelling of convection. 
An important difference is for example that 107 TPs are experienced
by the 6$\msun$ model by KL07 (see their tab.4), whereas only 26 by
our 6$\msun$ star.
This difference is due in part to the large luminosity determined by the
FST modelling, that favours a larger mass loss, and in part to the
different prescription for mass loss used by the two groups, since
the Bl\"ocker (1995) recipe predicts a more rapid increase in mass loss
with luminosity than the Vassiliadis \& Wood (1993) formula used by
KL07.

\section{Abundance trends expected in clusters of higher metallicity}

We can now compare the present results and those for smaller metallicity presented in
VD08. In the hypothesis that self--enrichment in GCs
is determined by the chemistry of the most massive AGBs, we can expect 
systematic trends with metallicity. The largest difference
is that metal rich cluster should show a milder oxygen depletion, 
limited to $\sim$0.4 dex, against the $\sim$0.8 dex for the Z=0.001 case.

\subsection{The sodium --- oxygen anticorrelation}

In the plane sodium vs. oxygen, the O--Na relationship at Z$\sim$0.004 clusters
is expected to have a higher slope than at Z$\sim$0.001, 
as the models with Z=0.004 show 
a sligthly higher enhancement of sodium, $\sim$[$^{23}$Na/Fe]=0.7 and a much lower
depletion of oxygen. Note that this prediction is independent of the
uncertainties related to the Ne-Na-Al proton capture cross sections, which,
on the other hand, limit considerably the predictive power concerning 
the sodium and aluminum enhancement shown by the ejecta of AGBs
(Izzard et al. 2007; VD08). 

\begin{figure*}
\includegraphics[width=.48\textwidth]{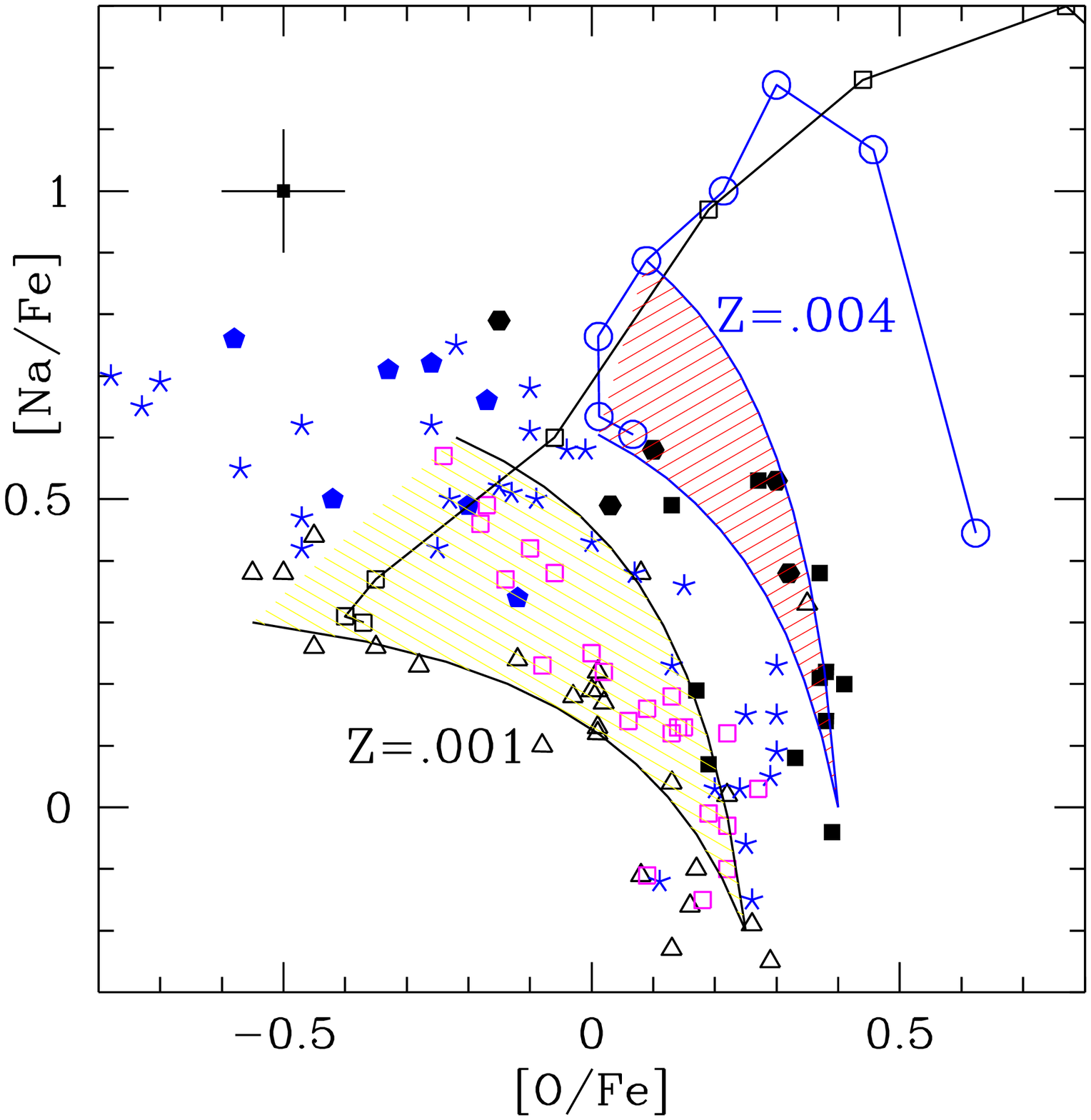}
\includegraphics[width=.48\textwidth]{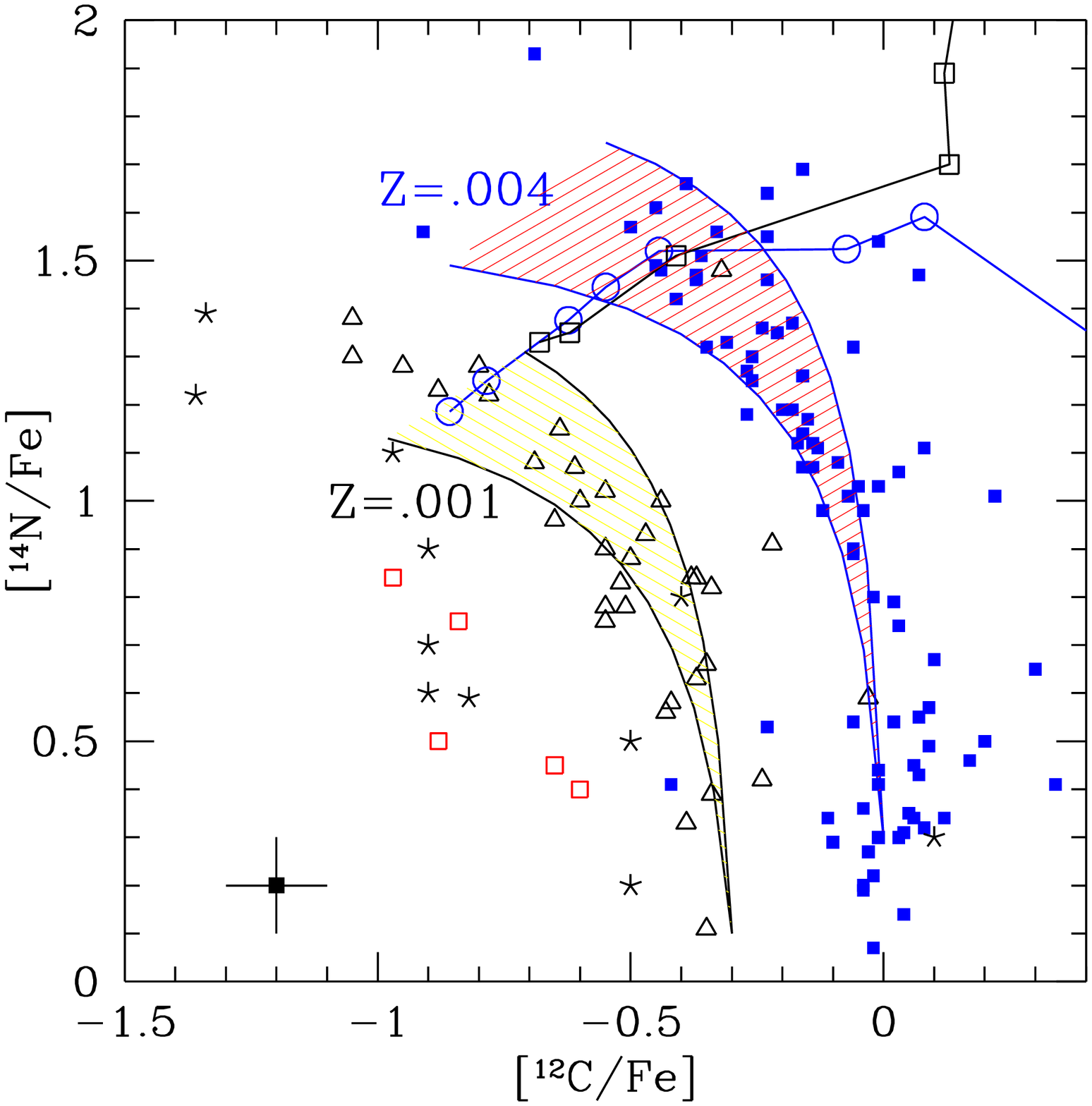}
\vspace{-50pt}
\caption{We show the Na--O (left) and N--C (right) anticorrelations for several GCs, 
in two main sets representing either metallicity
close to Z=0.004 or to Z=0.001. In the left figure, the points represent abundances of 
sodium and oxygen from different sources. Low Z clusters: open triangles: 
M5 giants from Ivans et al. 2001, stars (M13) and
open squares (M3) both from Sneden et al. 2004. High Z clusters: full squares:
M71 from Sneden et al. 1994; full hexagons:
NGC~6441 from Gratton et al. 2007; full pentagons: NGC~6388 from Carretta et al. 2007;
In the right figure the full dots are the data for the main sequence in 47~Tuc (high Z), 
from Briley et al. 2004. Data for M5 (open triangles) are from Cohen et al. 2002, 
for M13 (stars) and M3 (open squares) from Cohen \& Melendez 2005. 
A conservative error bar is also shown in the two panels. In both figures,
the connected open circles are the yields from Table 1 for Z=0.004, while the 
connected open squares are the yields from VD08 for Z=0.001. 
The dashed areas represent the abundances allowed for self--enriched stars,
for the two metallicites, if their chemistry results from matter lost from 
AGBs of masses down to 5\Msun, diluted 
in different percentages with pristine matter having the chemical 
composition of the vertex of the shaded area (abundance of the first generation 
stars) chosen as representative of the most 
normal abundances of the stars in the two sets of clusters 
(see discussion in the text concerning the yield normalization).
}
 \label{oxna}
\end{figure*}

In the left side of Fig.\ref{oxna} we plot the sodium versus
oxygen abundances in several GCs of metallicity close to Z=0.001 or to
Z=0.004. Notice that the relative errors in the points location
is at least $\sim$0.1~dex in the abundances, and that it is not easy to
judge possible additional systematic differences between abundance determination by
different researchers. The connected squares represent the abundances
expected in the AGB ejecta of masses from 3.5\msun\ (upper square on the right) 
to 6.3\msun\ for metallicity Z=0.001, from VD08.
The connected big open circles represent the abundances from the 
present models (Table 1) from 3 to 6.5\Msun. 

The comparison clusters are M5 \citep[open triangles, from][]{ivans2001}
and M3 \citep[open squares, from][]{sneden2004}, whose trend of abundances are well 
reproduced, and M13 \citep[stars, from][]{sneden2004}. This latter cluster possibly
requires a different normalization, and also shows  some very 
extreme oxygen abundances, below [O/Fe]=-0.4. The latter stars, however, 
are all luminous giants in all GCs so far examined \citep{carretta2006} and their
abundances could reflect in situ mixing, possibly due to the lower mean molecular
weigth barrier of giants starting with extreme helium abundances Y$>$0.35 
\citep{dantona2007}.
The three clusters of high metallicity included in the data are M71 
\citep[full squares, from][]{sneden1994}, 
NGC 6388 \citep[full pentagons, from][]{carretta2007} and NGC 6441 
\citep[full hexagons, from][]{gratton2007}. 

The stars having high oxygen and low sodium can be interpreted as the first generation
stars, born with the abundances of the pristine material from which the cluster formed.
The stars having high sodium and possibly also low oxygen are the self--enriched
stars. As the theoretical yields do not follow the trends defined by the stars, we
must interpret the data of self--enriched stars as the result of mixing of matter
having the yields of the AGBs with different relative amounts of pristine matter.
This dilution model can be applied to the yields of all masses above $\sim$5\Msun,
for which the TDU and the lower temperatures of the HBB does not lead to
too high sodium values. Among the smallest masses we consider, the 
total C+N+O abundance is increased due to the action of the third dredge up, but
not too much. The approximate constancy of total CNO is
another requirement of the observations \citep{ivans1999}, but actually, the
analysis of the CNO abundances in a few clusters by \citep{carrettacno2005} 
show that possibly some 
additional carbon from triple-$\alpha$\ processing may be needed, consistent 
with our choice to include stars in which some effects of the TDU
are already important.  

Let us then that stars from the upper mass computed down to 5\msun\ contribute to 
self-enrichment for Z=0.001 and Z=0.004. If the ejected matter and the
original matter from which the GC was born are mixed in different percentages,
the resulting abundances of the second
generation stars will be contained in the dashed areas. Of course, part of --or also 
most of-- the points close to the vertex of the area in fact may be stars 
directly formed in the first stellar generation. 

In order to apply the dilution model, we have to assume an oxygen and sodium abundance
for the original matter. For M71, Fig. \ref{oxna} shows that we can reasonably assume 
that the chemistry of the first generation stars (and of the pristine gas) is 
[O/Fe]$\sim$0.4 and [Na/Fe]$\sim$0.0. As these are the initial abundances in our models,
for Z=0.004 we can assume that the yield abundances of the matter ejected from
the AGBs are those directly given in Table 1 and in the connected circles in the
figure. For the low metallicity clusters in the figure, the zero point of 
oxygen abundance is closer to [O/Fe]$\sim$0.25
and [Na/Fe]$\sim -$0.2. So we have scaled down to these values 
the oxygen and sodium yields of AGBs, with respect to the plotted values 
(concerning the validity of this approach, see the discussions 
in \cite{ventura2005b} and VD08). 

We see that the dilution model does a good job in describing the fact that the higher
metallicity stars, those in M71, reach larger sodium abundances that 
those in the metal poor clusters, as our models predict. The data
however are very scarce indeed, and require further confirmation. In addition, 
{\it all} the data points for NGC 6388 are at low oxygen -- high sodium: 
where is the 
``standard" first population at high oxygen -- low sodium? Also in this case, further
data are needed. Although the oxygen abundances in 
NGC 6388 are not as extreme as in M13, 
we may again interpret these stars as giants suffering extra--mixing as proposed 
by \cite{dantona2007}: this cluster, indeed, as its twin NGC 6441, has a horizontal
branch morphology which implies that $\sim 20$\% of its stars have Y$>$0.35 (and
another 40\% has 0.35$<$Y$<$0.27) \citep{caloidantona2007, dantonacaloi2007}.

In summary, Figure \ref{oxna} shows that our models are in agreement
with the observational data, if the range of AGB stars that contributes to
self enrichment is limited down to $\sim$5\Msun. However,
the observations must be extended to larger samples before we can claim consistency. 

\subsection{Carbon and nitrogen}
Another comparison between the element trends expected for different metallicities  
can be done on the carbon and nitrogen abundances. 
Unfortunately, only some of the clusters examined for the Na--O anticorrelation have 
also data for C and N. In the right part of Figure \ref{oxna}
we show the N versus C abundances for the low metallicity 
clusters M5 \citep{cohen2002}, M3 and M13 \citep{cohen-melendez2005}
(same symbols as in the Na--O figure on the left). As representative of 
a metallicity closer to the new models, Z=0.004, we plot the data for stars close
to the main sequence in 47 Tuc, by \cite{briley2004}; as discussed in detail by
\cite{briley2004}, also M~71 has a very similar C--N behaviour. 
The model yields from Table 1 and from VD08, reported in Fig. \ref{oxna}, 
show that the most massive models of Z=0.004 (open circles) have {\it lower} $^{12}$C
and $^{14}$N than the most massive models of Z=0.001 (open squares). 
This is easily understood, considering that the 
ON processing is stronger at lower metallicity, and 
therefore a larger amount of nitrogen results from the full CNO cycling. 
{\it In principle this result looks at variance with the observations}, 
as the 47 Tuc (high Z) data show more Nitrogen than the M5 (low Z) data. Nevertheless,
the abundances derived are probably affected by a zero point uncertainty: e.g., the few 
data by \cite{carrettacno2005}) for 47~Tuc have [C/Fe] $\sim$0.3dex 
smaller than \cite{briley2004}. 
Obviously, some scaling must be applied to the theoretical yields, to
take into account the differences in initial abundance of the
pristine matter. We already used this
procedure for the sodium vs. oxygen yields, assuming that the initial values  
[O/Fe]=0.4 and [Na/Fe]=0.0 at Z=0.001 were reduced to values [O/Fe]=0.25 and 
[Na/Fe]=--0.2, but in the case of C and N this operation is a bit more
tricky, because either the ``primordial" carbon 
abundance of M5 is really smaller than the [C/Fe]$\sim$0 of the models (and also
of the data for 47~Tuc), or the difference may be due to
a difference in absolute scale.  
Lower initial carbon abundances in M5 implies that less nitrogen
is produced by the CN cycle, so we must also reduce the theoretical yield of Nitrogen
by the same logarithmic amount. But if the difference is only a zero point problem, 
on the contrary, the strength of the CN band leads to derive a larger [N/Fe] abundance
(in fact, this could in principle give origin to an anticorrelation in the results
---\cite[][]{cohen2002, briley2004}). 

As the stars in 47 Tuc are easily divided in two main groups:
a population where carbon and nitrogen are practically untouched, clustering
at [C/Fe]$\sim$0, [N/Fe]$\sim$0.3, and a
group of stars which define a clear C--N anticorrelation, we consider
possible pollution from the range of masses between 5 and 6.5, as in the
sodium vs. oxygen case, and define the dashed upper area. 
The data points are well contained in the region defined by the theoretical models.
For Z=0.001 we fit the M5 data, by scaling down the carbon abundance and the total nitrogen
abundance by a factor two. In this way, we are assuming 
that the initial C abundance of M5 is really smaller by --0.3~dex than the solar 
scaled value assumed in the models, and not that we are in the presence of a 
possible zero point uncertainty.

The comparison with M5 and 47 Tuc provides a reasonable agreement with the
theoretical expectations, although the comparison remains ambiguous, as outlined above.
The data for M13 and M3 are scarce and it is not clear which is the
starting abundance, so no comparison with the models is attempted.

\subsection{Aluminum}
In the metal rich GCs, the
enhancement of Aluminum should be limited to [$^{27}$Al/Fe]=0.6, 
to be compared to [$^{27}$Al/Fe]$\simeq$1 in the Z=0.001 case described in VD08.
Thus we expect again an O-Al anticorrelation in
higher metallicity GCs, with the same trend of the O-Al anticorrelation
in lower metallicity clusters, but extended to less extreme abundances of O and Al.
\begin{figure}
\includegraphics[width=.48\textwidth]{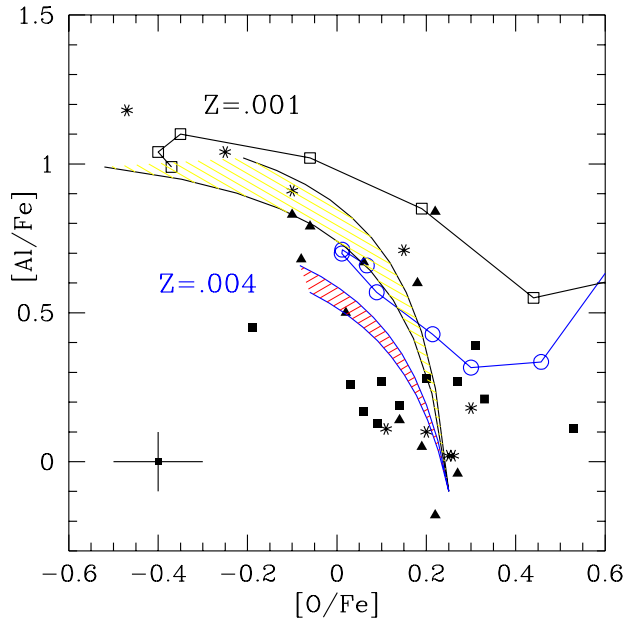}
\vspace{-50pt}
\caption{We show the Al--O anticorrelation for M3 (full triangles), 
 and M13 (asterisks) from Sneden et al.(2004), and for
 M71 (full squares) by Ramirez \& Cohen (2002). The yields for Z=0.001
 (open squares) and for Z=0.004 (open circles) are also plotted.
 The two dashed regions represent the abundances allowed for
 self-enriched stars, as in Fig.~\ref{oxna}.}
\label{allum}
\end{figure}
We show in Fig.~\ref{allum} the Al--O data for M13 and M3 from Sneden 
et al. (2004), for M71 from Ramirez \& Cohen (2002), 
and the yields from the models of 
Z=0.001 and Z=0.004. We then perform the same analysis
made for Na--O and N--C in Fig.~\ref{oxna}. We assume that the abundances derive from 
different mixtures of matter ejected by the stars in the range 5--6.3\Msun (Z=0.001) or
5--6.5\Msun (Z=0.004) with pristine matter having abundance [O/Fe]=0.25 and [Al/Fe]=0.1.
The dashed areas show the difference between the two metallicities predictions.
We also see that the M3 and M13 data are in very good agreement with the expectations for 
Z=0.001. Notice that, while the models predict that the Na--O and N--C anticorrelations 
should show large spread, {\it the Al--O anticorrelation should be much more tight} as
the oxygen and aluminum yields of the range of masses allowed are very close. The data
for M~71 have in fact much smaller Al variations than the low Z clusters data, but a
stringent comparison would need many more data.

\section{Conclusions}
We present new AGB models with metallicity Z=0.004, to test the hypothesis of AGB
self-enrichment in high metallicity GCs and compare the results with previous
models having Z=0.001 (VD08). We find that the main physical
features of the AGB evolution of the intermediate mass stars do not change by
increasing the metallicity, the main
effects being only a shift towards higher masses of the mass-luminosity
relationship; models with the same core mass and different Z follow 
very similar evolution, in
terms of the temperature reached at the bottom of their convective envelope.

The theoretical yields of the most massive AGB models are in agreement 
with the aluminum, sodium and oxygen abundances of the most anomalous stars 
(those showing the strongest depletion of oxygen ---and excluding the giant in which
the very low O should be attributed to deep extramixing)
in the few high metallicity GCs so far examined; a reasonable
dilution scheme gives results consistent with the O--Na and C--N anticorrelations. 
Yet, other confirmation is needed, since the data currently available involve
only few stars per cluster. 

One robust prediction of this investigation, which is independent of all
the uncertainties associated to the proton capture cross sections by Ne-Na-Al
nuclei, is that with increasing metallicity we expect a higher slope of
the O-Na anticorrelation, and a similar slope (though less extended) of
the O-Al trend.

\end{document}